\documentclass[conference,a4paper]{IEEEtran}
\IEEEoverridecommandlockouts

\usepackage{cite, graphicx, algorithmic}
\usepackage{amsmath, amssymb, amsfonts}
\usepackage{textcomp, xcolor, siunitx}
\usepackage{svg, textcomp, subfigure, url}

\usepackage{setspace}

\begin{document}

\title{A Tsetlin Machine Image Classification\\ Accelerator on a Flexible Substrate
\thanks{This work was supported by EPSRC EP/X039943/1 UKRI-RCN: Exploiting the dynamics of self-timed machine learning hardware (ESTEEM) project and by I-UK project Mignon Automated Hardware Generation for Embedded Machine Learning Applications (MARGE). Special thanks to Dr. Yujin Zheng for her help with this paper.}
}

\author{
\IEEEauthorblockN{Yushu Qin\IEEEauthorrefmark{1}, Marcos L. L. Sartori\IEEEauthorrefmark{1}\IEEEauthorrefmark{2}, Shengyu Duan\IEEEauthorrefmark{1}, Emre Ozer\IEEEauthorrefmark{1}\IEEEauthorrefmark{2}, Rishad Shafik\IEEEauthorrefmark{1}, Alex Yakovlev\IEEEauthorrefmark{1}}
\IEEEauthorblockA{\IEEEauthorrefmark{1} Microsystems group, School of Engineering, Newcastle University \\ Newcastle Upon Tyne, UK}
\IEEEauthorblockA{\IEEEauthorrefmark{2} Pragmatic Semiconductor Ltd, Cambridge, UK}
\IEEEauthorblockA{\{y.qin20,marcos.sartori,shengyu.duan,rishad.shafik,alex.yakovlev\}@newcastle.ac.uk,\ eozer@pragmaticsemi.com}
}

\maketitle
\begin{abstract}
This paper introduces the first implementation of digital Tsetlin Machines (TMs) on flexible integrated circuit (FlexIC) using Pragmatic’s 600nm IGZO-based FlexIC technology. TMs, known for their energy efficiency, interpretability, and suitability for edge computing, have previously been limited by the rigidity of conventional silicon-based chips. We develop two TM inference models as FlexICs: one achieving 98.5\% accuracy using 6800 NAND2 equivalent logic gates with an area of 8×8 mm\textsuperscript{2}, and a second more compact version achieving slightly lower prediction accuracy of 93\% but using only 1420 NAND2 equivalent gates with an area of 4×4 mm\textsuperscript{2}, both of which are custom-designed for an 8$\times$8-pixel handwritten digit recognition dataset. The paper demonstrates the feasibility of deploying flexible TM inference engines into wearable healthcare and edge computing applications.
\end{abstract}
\begin{IEEEkeywords}
Tsetlin Machine, Flexible integrated circuit (FlexIC), Machine learning, Image classification
\end{IEEEkeywords}
\section{Introduction}

Tsetlin Machine (TM)\cite{TM}, a logic-based machine learning algorithm, has recently gained substantial attention due to its interpretability, energy efficiency, and simplicity in hardware implementation. Unlike conventional neural networks\cite{mcculloch_NN_1943}, TMs rely on propositional logic rules and discrete state machines, making them particularly suitable for energy-constrained edge computing applications such as wearable health monitoring devices and IoT sensors. Early efforts have already demonstrated efficient hardware implementations on FPGA and ASIC platforms \cite{Adrian-TMhardware}. 

Advances in flexible integrated circuit (FlexIC) technology have expanded the possibilities of embedding computational intelligence directly into flexible substrates. For example, \textit{Pragmatic} develops a FlexIC technology based on fabricating Indium-gallium-zinc-oxide (IGZO)-based metal-oxide thin-film transistors on a polyimide substrate \cite{FlexIC}. FlexICs are ultra-thin, low-cost and physically flexible or bendable. They are suitable for wearable and biomedical applications where form factor, conformability and thinness are essential. Prior works on the feasibility of developing general-purpose processors \cite{biggs_flexARMChip_2021, ozer_RISCV_2024, flex6502} and low-complexity machine learning hardware accelerators \cite{flexml, flexbnn, tcc, ozer_malodour_2023, arrwnn} as FlexICs. 

Conventional silicon-based application specific integrated circuits (ASICs) used in current TM implementations typically lack mechanical flexibility, limiting their practicality in wearable and flexible electronics scenarios.
Although flexible electronics have matured in materials and fabrication techniques, their suitability for implementing TM accelerators has not yet been demonstrated.

In this paper, we design and implement the first TMs on flexible substrates aiming at edge computing applications. Two TMs are designed: 1) A full-scale model with approximately 6800 NAND2-equivalent logic gates, achieving an accuracy of 98.5\% on an 8$\times$8-pixel handwritten digit recognition dataset; 2) A more compact variant consisting of only 1420 NAND2-equivalent gates but achieving an accuracy of 93\%.

The rest of the paper is organised as follows: \emph{Section \ref{methodology}} describes the methodology of training the two TM models for MNIST dataset and the design and implementation of the two FlexICs for the two TM inference models. \emph{Section \ref{experiments}} presents the experimental results of the two TM FlexIC hardware. \emph{Section \ref{conc}} and \emph{Section \ref{future}} conclude the paper with some future work.

\section{Methodology}
\label{methodology}
This section describes the methodology of training the two TM models for a dataset of handwritten digit recognition \cite{optical_recognition_of_handwritten_digits_80} in software prior to the design.
Once training is complete, the inference models of the TMs are designed as hardware blocks by translating them into RTL code, which is then fed into the hardware synthesis and physical implementation EDA tools to develop the two inference TMs FlexICs.

\begin{figure*}[htb]
    \centering
    \includegraphics[width=.9\linewidth]{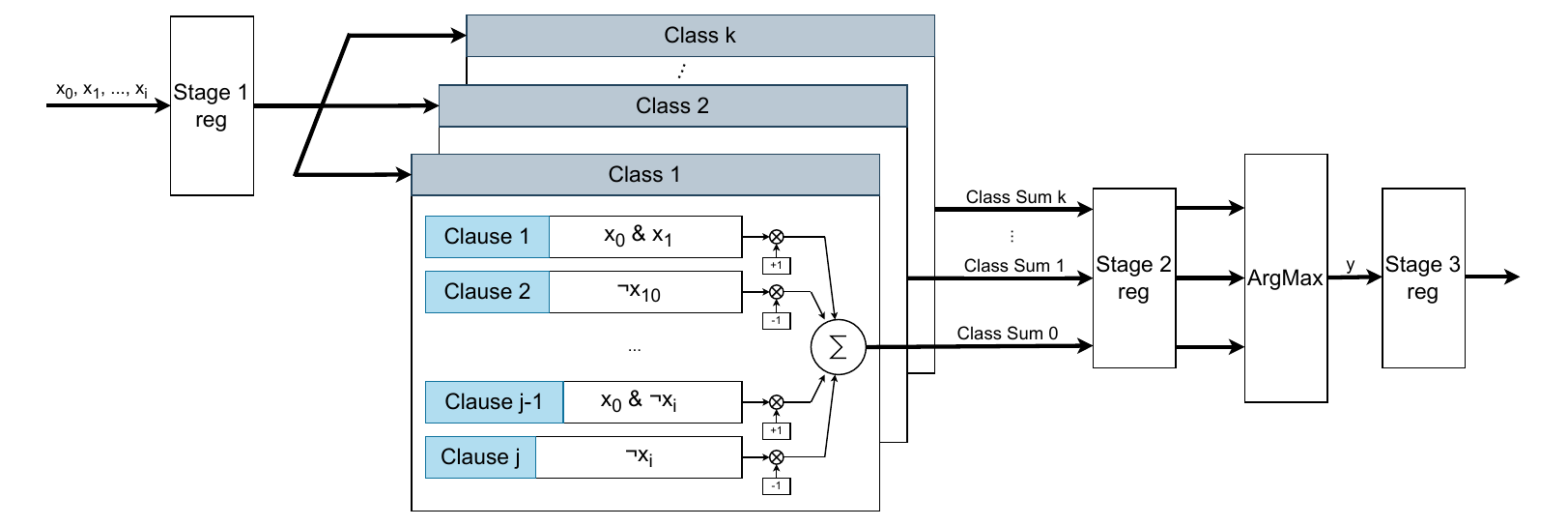}
    \caption{Hardware microarchitecture of the two TM inference models}
    \label{pipeline RTL}
\end{figure*}

\subsection{Model Generation}

Rather than training a single TM model by varying the number of epochs, we generate a series of TMs under identical small hyperparameters \(T\) and \(s\), exploiting the stochastic nature of the learning process to obtain diverse accuracy profiles.
The dataset used in this section is a handwritten digit recognition dataset consisting of 8$\times$8-bit input images and 10 output classes (digits 0–9) \cite{optical_recognition_of_handwritten_digits_80}.

Specifically, the hyperparameter $T$ reflects the model's confidence in distinguishing between classes. By increasing or decreasing $T$, more or less feedback is triggered during training, which in turn leads to the learning of more or less complex patterns.
Conversely, the parameter $s$ influences the inclusion probability of literals via Type I feedback. As $s$ increases, the model tends to form more complex decision patterns.
Therefore, the model complexity—and consequently, the implementation cost—can be controlled by appropriately tuning the parameters $T$ and $s$, while generally preserving similar classification accuracy \cite{tarasyuk2023systematic}.

First, we determine the appropriate number of training epochs \(E\) through a pilot study: A single TM (with \(T=10\), \(s=3.0\), 10 output classes and 100 clauses per class) was repeatedly trained, incrementally increasing the epoch count until the validation accuracy plateaued; the smallest epoch number beyond which further increases yielded negligible improvement is selected as \(E\). This procedure allow us to fix \(E = 100\) for all subsequent experiments.

Under this configuration (\(T=10\), \(s=3.0\), \(E=100\), 10 classes, 100 clauses per class), each TM instantiation begins with a different random seed, which influences the weight updates of the clause during learning. As a result, despite sharing the same hyperparameters, individual models converge to decision boundaries of varying quality and complexity.

To systematically explore this variability, we generate \(N = 500\) independent TM models. For each model \(M_i\), we record its validation accuracy \(A_i\) after completion of the fixed 100 epochs. Across all 500 models, we observe an average accuracy of \(98\%\), with the highest accuracy reaching \(98.5\%\). Models exhibiting the highest accuracy values are then selected for further analysis, while the remaining models provide insights into the spread of performance under fixed hyperparameter constraints.

By using the same method but changing the number of clauses in the TM as well as the hyperparameter \(T\) and \(s\) to a smaller value, a compact model can be generated with simpler clause expression but slightly lower accuracy, here is the detailed comparison of the compact model and the full-scale one:
\begin{itemize}
  \item Full-scale model: 10 classes, 100 clauses per class, using 9200 Tsetlin Automata (TA) included in the clause expression; achieves \(98.5\%\) accuracy with \(T=10\), \(s=3.0\).
  \item Compact model: 10 classes, 20 clauses per class, pruned to 559 TA in clause expressions; uses \(T=5\), \(s=1.8\) to maintain competitive performance with substantially fewer clauses, average accuracy is \(89\%\) while the highest is \(93\%\).
\end{itemize}

By leveraging many TMs trained with a small hyperparameter footprint and selectively scaling clause counts, we maintain low memory and computational requirements per baseline model while harnessing model diversity and targeted configuration to achieve both efficiency and state-of-the-art accuracies.

\subsection{Design}
The two trained TM inference models are translated into hardware description to implement them as TM inference engines in FlexIC technology. 

The hardware microarchitecture of the TM models is organised into a four-stage pipeline to optimise throughput and manage resource utilisation.
\textbf{Fig.~\ref{pipeline RTL}} illustrates the pipelined design of the TM inference models.
This model receives a 64-bit input vector \(x\) and outputs a 4-bit array \(y\) which indicates the index of the winning class. In the first stage, all input bits are immediately captured by a series of registers directly connected to the chip's I/O pins.
This first register layer isolates the internal logic from external delays, aligning the input data and avoiding spurious transitions.

The second stage comprises the core combinational logic of the TM.
The registered inputs flow into the clause calculation block, where all propositional clauses are evaluated in parallel.
The clause votes produced by this block are then fed into the class summation unit, a tree of adders that aggregates votes for each output class.
Because this combined clause evaluation and summation path is relatively deep, its outputs are captured in a dedicated set of pipeline registers.
These mid-stage registers isolate timing, allowing the design to achieve a higher clock frequency.

Lastly, the resulting counted class votes enter a purely combinational ArgMax unit, which compares all class sums and selects the index of the highest vote count.
The resulting class index is then buffered in a final register stage before the output pins are driven.
Registering the ArgMax result simplifies the timing closure for any external circuits that consume the classification result.

In summary, the TM is implemented in a four-stage architecture encompassing (i) input literal capture; (ii) clause computation and summation; (iii) class comparison; and (iv) output buffering. The TM hardware description is written in Verilog and synthesised using \textit{Cadence Genus}.

\subsection{Physical Implementation}

\begin{figure}[htb]
    \centering
    \includegraphics[width=.9\linewidth]{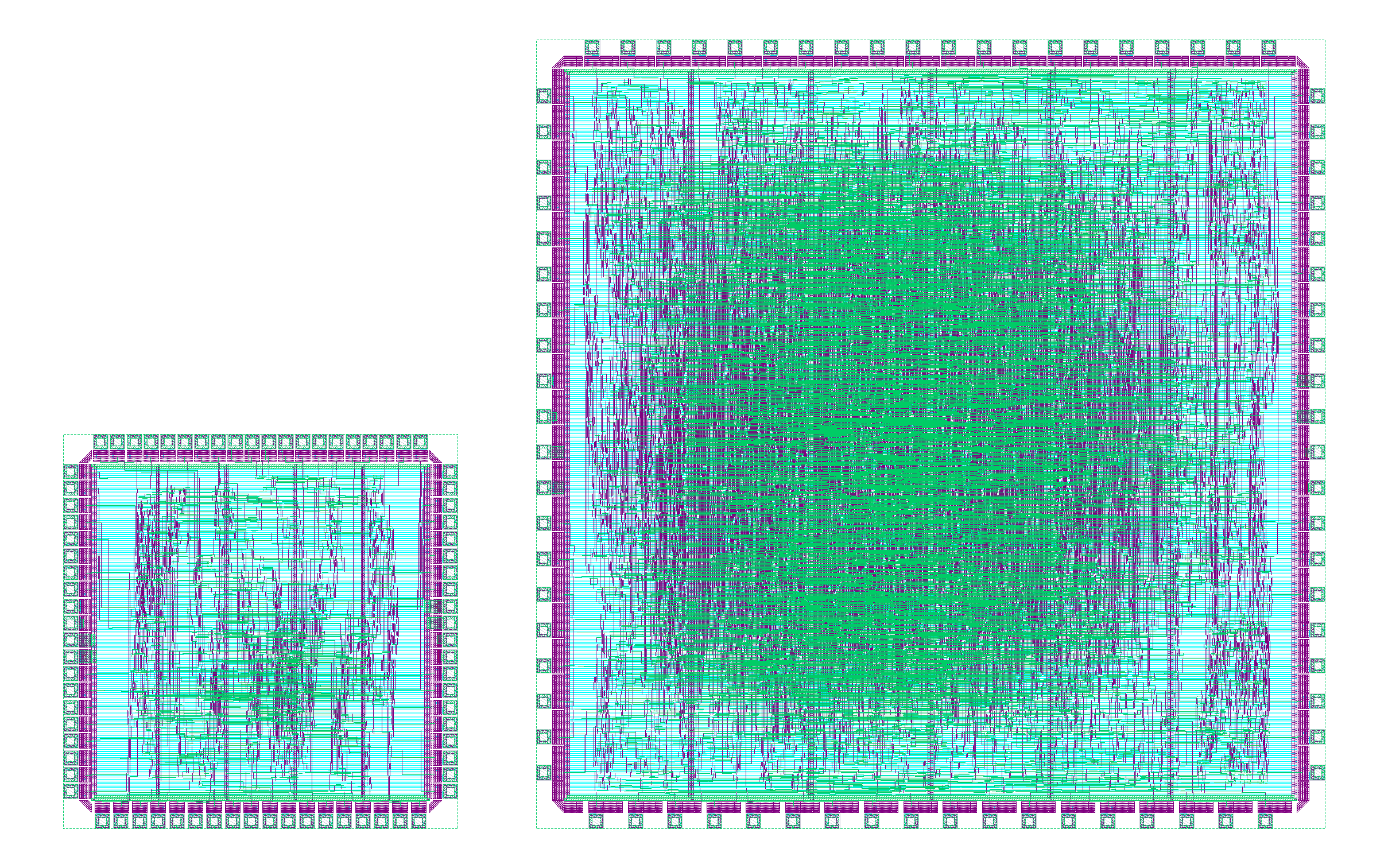}
    \caption{Layouts of the compact and full-scale TM FlexICs}
    \label{layout}
\end{figure}

The two TM inference models are implemented in \textit{Pragmatic}’s 600nm IGZO-based FlexIC process \cite{FlexIC} using a standard-cell library for core logic and a fully custom I/O pad library. Two die sizes are targeted 8 mm × 8 mm for the full-capacity model and 4 mm × 4 mm for the compact model to explore the trade-off between compute density and form factor on a bendable substrate. Physical implementation including floorplanning, placement, and routing is performed using\textit{Cadence Innovus}.

To prevent routing congestion under the two-layer constraint, we enforce a maximum cell placement density of approximately 10\% within the core’s power-ring boundary. During floorplanning, a continuous power ring of VDD and GND surrounds the core region; within this ring, placement directives in \textit{Innovus} (e.g., utilisation constraints) ensure that standard-cell clusters never exceed the 10\% density target. VDD and GND networks are realised through this outer ring, supplemented by on-chip metal-stripe rails and “special route” segments that tie the ring and stripes together. Each supply (VDD and GND) is driven by two dedicated power pads to minimise IR drop and support the high switching currents. 

Signal routing, power planning, and timing closure are all handled by \textit{Innovus}’s global and detailed router. Despite the limited metal stack, careful density control and congestion-driven optimisations allows the design to meet timing requirements without resorting to additional metal layers. Post-route verification — both Design Rule Check (DRC) and  Layout Versus Schematic (LVS) — passed cleanly, confirming that the layout adheres to the foundry’s design rules and matches the schematic. The resulting FlexICs demonstrate that, even with minimal routing resources, a disciplined placement strategy and EDA–tool–driven optimisations can deliver robust, manufacturable AI accelerators on flexible substrates. The final layouts of the full-scale and compact TM FlexICs are illustrated as \textbf{Fig. \ref{layout}}.

\section{Experimental Results}
\label{experiments}
\subsection{Simulation}

\begin{figure}[htb]
    \centering
    \includegraphics[width=\linewidth]{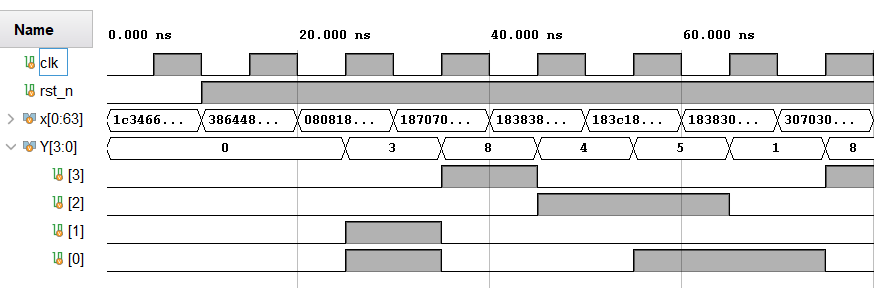}
    \caption{RTL simulation of the compact design}
    \label{RTL Sim}
\end{figure}

RTL verification is performed using a fully standalone SystemVerilog testbench that instantiates the TM DUT and drives its inputs with real-world stimulus vectors. The simulation results are shown in \textbf{Fig. \ref{RTL Sim}} where 4 signals are displayed: 1) System clock signal \(clk\), 2) Synchronous negative activating reset signal \(rst\_n\), 3) A 64-bit input signal \(x\), and 4) A 4-bit output signal \(y\). All input signals change at the falling edge of the clock signal to ensure that the system can recognise them correctly at the rising edge of each clock cycle. The results demonstrate that upon presenting an input vector, it correctly produces the ArgMax index exactly three clock cycles later, confirming its functional correctness.

Post-layout simulation with parasitic resistance and capacitance is performed after importing the layout into \textit{Cadence Virtuoso}. Prior to exporting parasitic data from the physical view, DRC and LVS must be performed using \textit{Siemens Calibre} to ensure that the layout complies with foundry design rules and accurately matches the intended schematic netlist. This verification step is critical for reliable parasitic extraction and subsequent timing and power analysis.

\begin{figure}[htb]
    \centering
    \includegraphics[width=\linewidth]{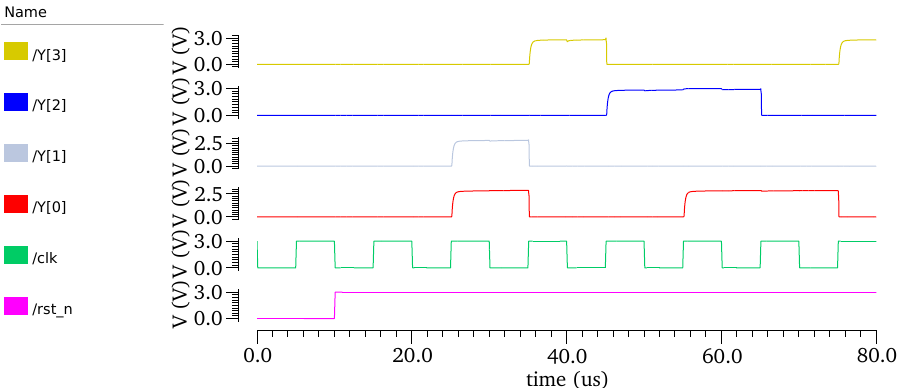}
    \caption{Parasitic simulation of the compact design}
    \label{PEX Sim}
\end{figure}

By using the same testbench setup from the RTL simulation performed above, the output waveform of the parasitic simulation is captured as \textbf{Fig. \ref{PEX Sim}} where it displays six signals: the system clock signal (clk), the synchronous negative activating reset signal (rst\_n), and a 4-bit output signal (y[3:0]). All output transitions exhibit a worst-case propagation delay of approximately 1.5\textmu s, which is well within the 10\textmu s period of the 100 KHz clock. This confirms that the design meets its timing requirements and operates reliably at the target frequency.

\subsection{Performance, Power, and Area (PPA) Analysis}

In this section, we present a comprehensive PPA analysis for the digital TM implementation under a wide range of clock‐period constraints. All measurements are obtained after synthesis, and gate counts are  in NAND2 equivalents. 



\begin{figure}
    \centering
    \subfigure[Full Model]{
    \includegraphics[width=.9\linewidth]{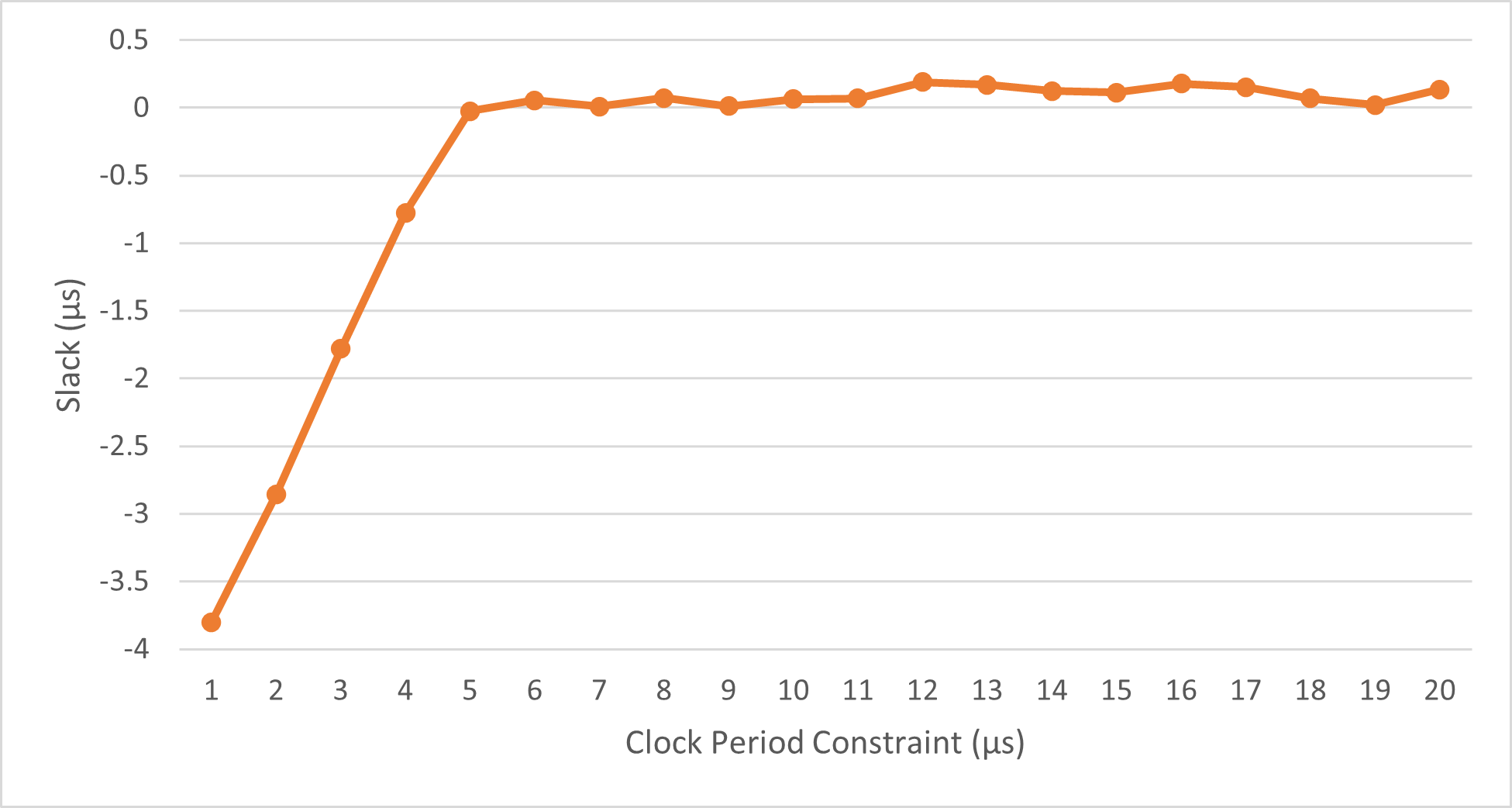}
    \label{Timing Full Model}}
    \quad
    \subfigure[Compact Model]{
    \includegraphics[width=.9\linewidth]{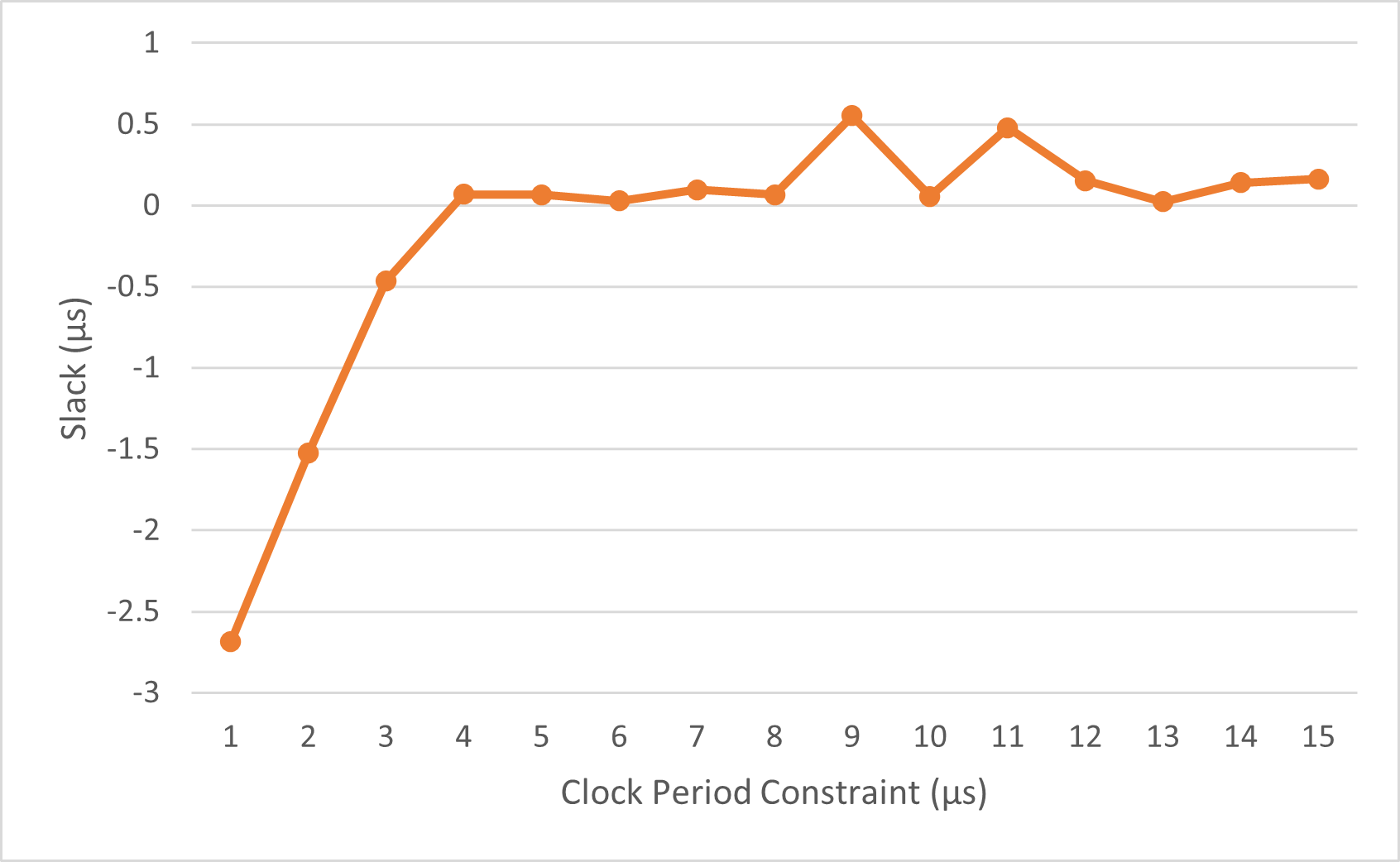}
    \label{Timing Comp Model}}
    \caption{Worst timing slack figures.}
    \label{WNS}
\end{figure}

\textbf{Fig. ~\ref{WNS}} depicts the worst timing slack for a range of synthesis of both designs subjected to a sweep of the clock period constraint.
It establishes the limits of the design space for power-performance-area (PPA) trade-off exploration.
It is possible to observe that the full model presents acceptable slack figures on clock constraint from \SI{5}{\micro\second} onwards, whereas the compact model fails at constraints below \SI{4}{\micro\second}.
After synthesis, negative-slack circuits are omitted from PPA analysis.

\begin{figure}[htb]
    \centering
    \subfigure[Full Model PPA Curve]{
    \includegraphics[width=.9\linewidth]{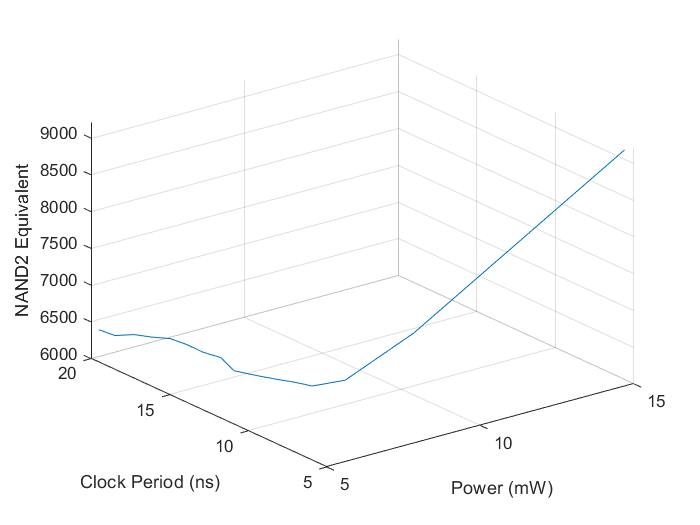}
    \label{PPA Full Model}}
    \quad
    \subfigure[Compact Model PPA Curve]{
    \includegraphics[width=.9\linewidth]{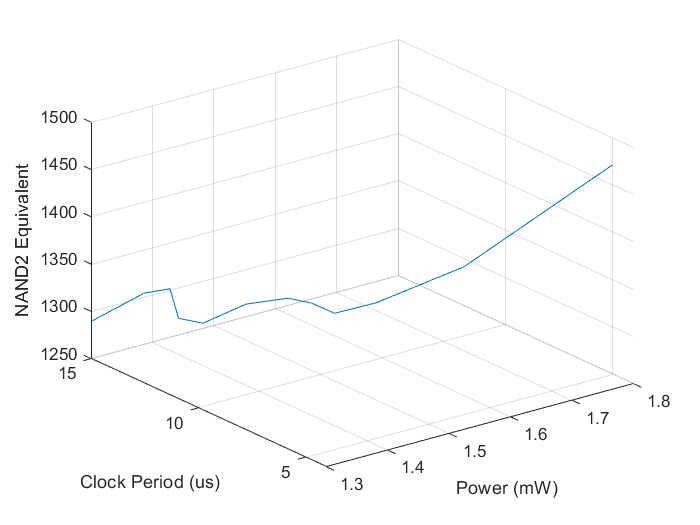}
    \label{PPA Comp Model}}
    \caption{PPA Curve for Both Models}
    \label{PPA}
\end{figure}

To visualise the multi-dimensional trade-offs, \textbf{Fig.~\ref{PPA}} presents three-dimensional curves for the Full and Compact TM models, respectively. In these plots, the axes represent clock period, power, and area (represented as gate count).
Every point along a curve represents a synthesis result for a particular clock period constraint.
Moving from the more relaxed to the higher performance designs; the least constrained designs present smaller power and area figures, but as the clock period is further constrained, Genus produces bigger and more power demanding circuits.

However when analysing the compact model closely, the curve shows at the most strict achievable clock period constraint of \SI{4}{\micro\second}, the design area is about 1480 equivalent-NAND2 gates with a power dissipation of 1.8 mW.
But, as the clock period constrains relaxes, this curve descend. At the \SI{15}{\micro\second} clock period constraint, the compact design requires only 1290 gates and about 1.3 mW.
A reduction of 27\% in power and 12\% in area for a 73\% reduction in clock frequency.
This shows that the design is very inelastic regarding the operating frequency.
The full-scale model (\textbf{Fig. \ref{PPA Full Model}}) exhibits a similar trade-off space, except for the tightest constraint where the clock period constrains approaches the design limit, presenting a slightly (but recoverable) negative slack.

\begin{figure}[htb]
    \centering
    \subfigure[Full Model]{
    \includegraphics[width=.9\linewidth]{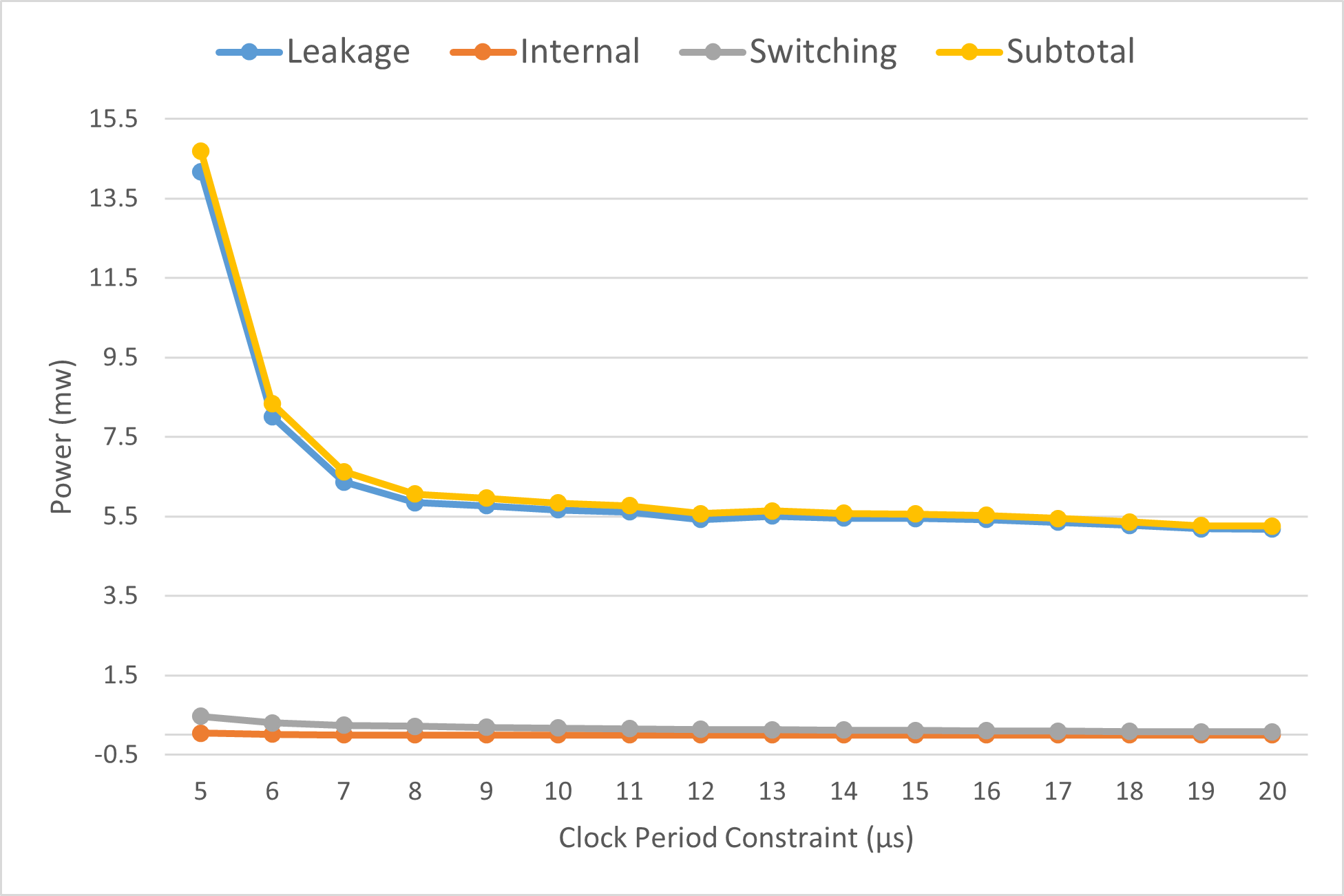}
    \label{Power Full Model Normal}}
    \subfigure[Compact Model]{
    \includegraphics[width=.9\linewidth]{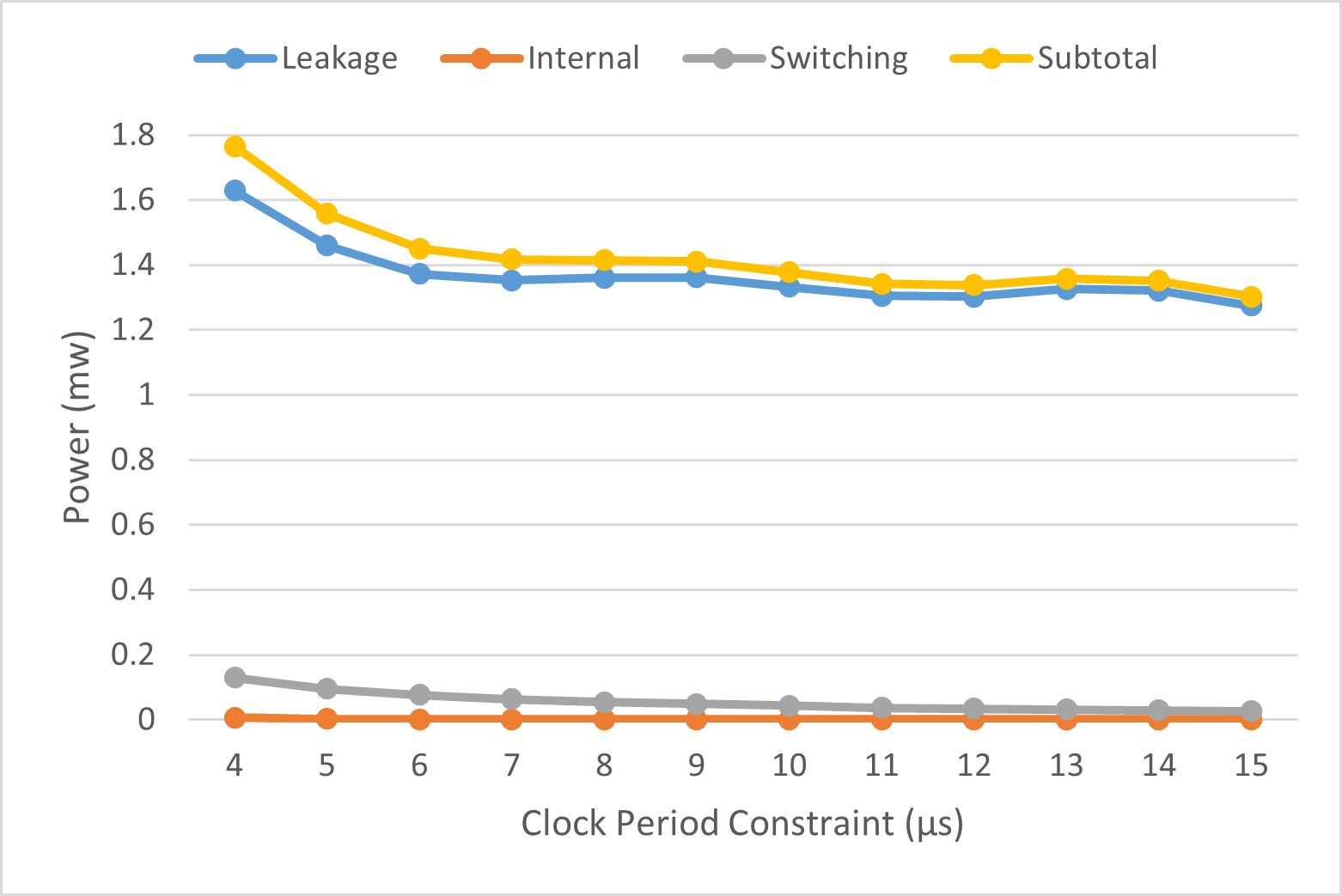}
    \label{Power Comp Model Normal}}
    \caption{Power Consumption for both TM Models}
    \label{Power}
\end{figure}

\textbf{Fig.~\ref{Power}} shows the power results broken down into its components.
Power consumption is mainly static, which
stems from the pull-up resistor used in unipolar logic gates of the FlexIC technology. This draws a continuous current through the logic gates when the output is driven low.

\begin{figure}[htb]
    \centering
    \subfigure[Full Model]{
    \includegraphics[width=.9\linewidth]{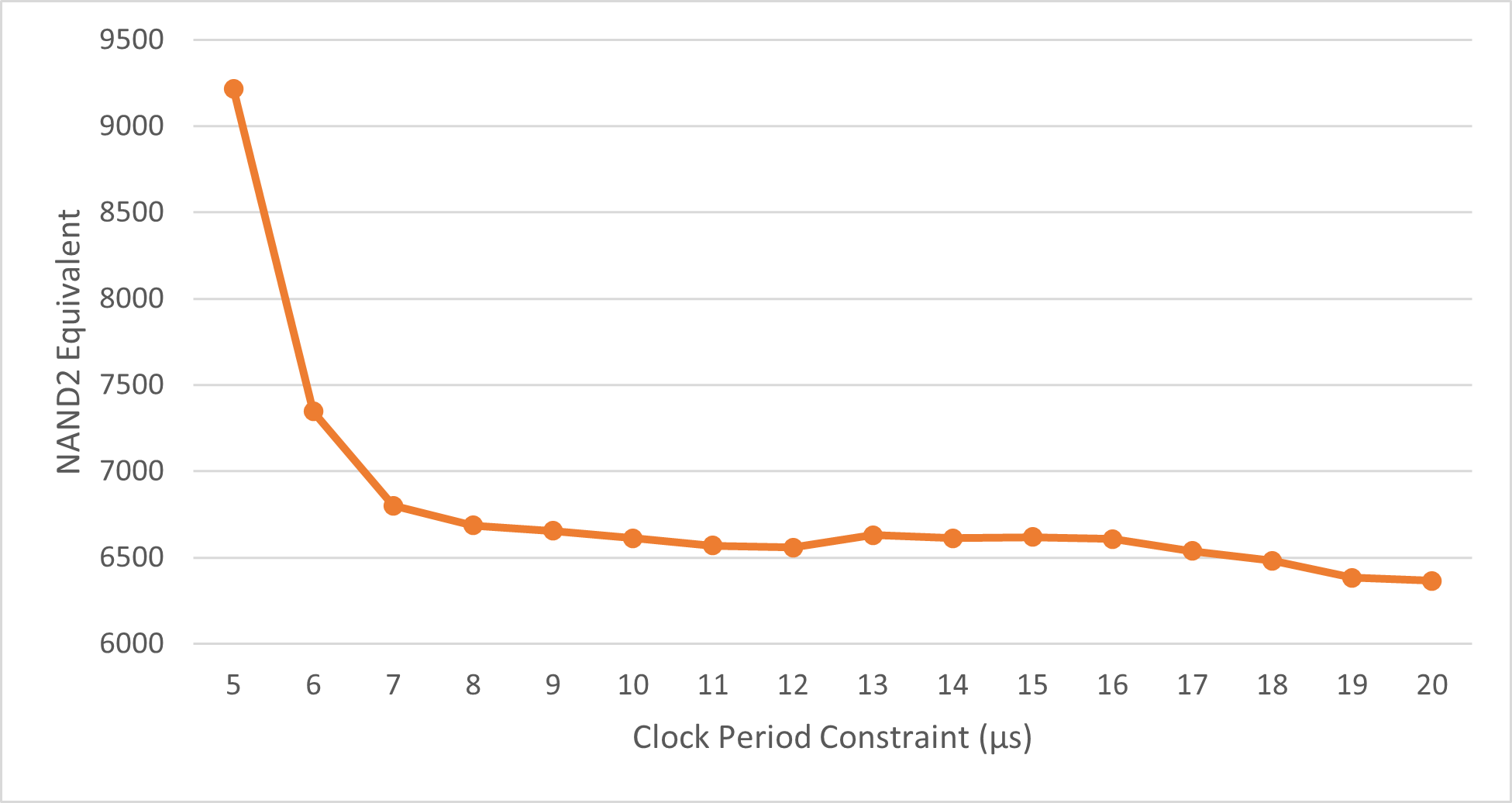}
    \label{NAND2 Full Model}}
    \quad
    \subfigure[Compact Model]{
    \includegraphics[width=.9\linewidth]{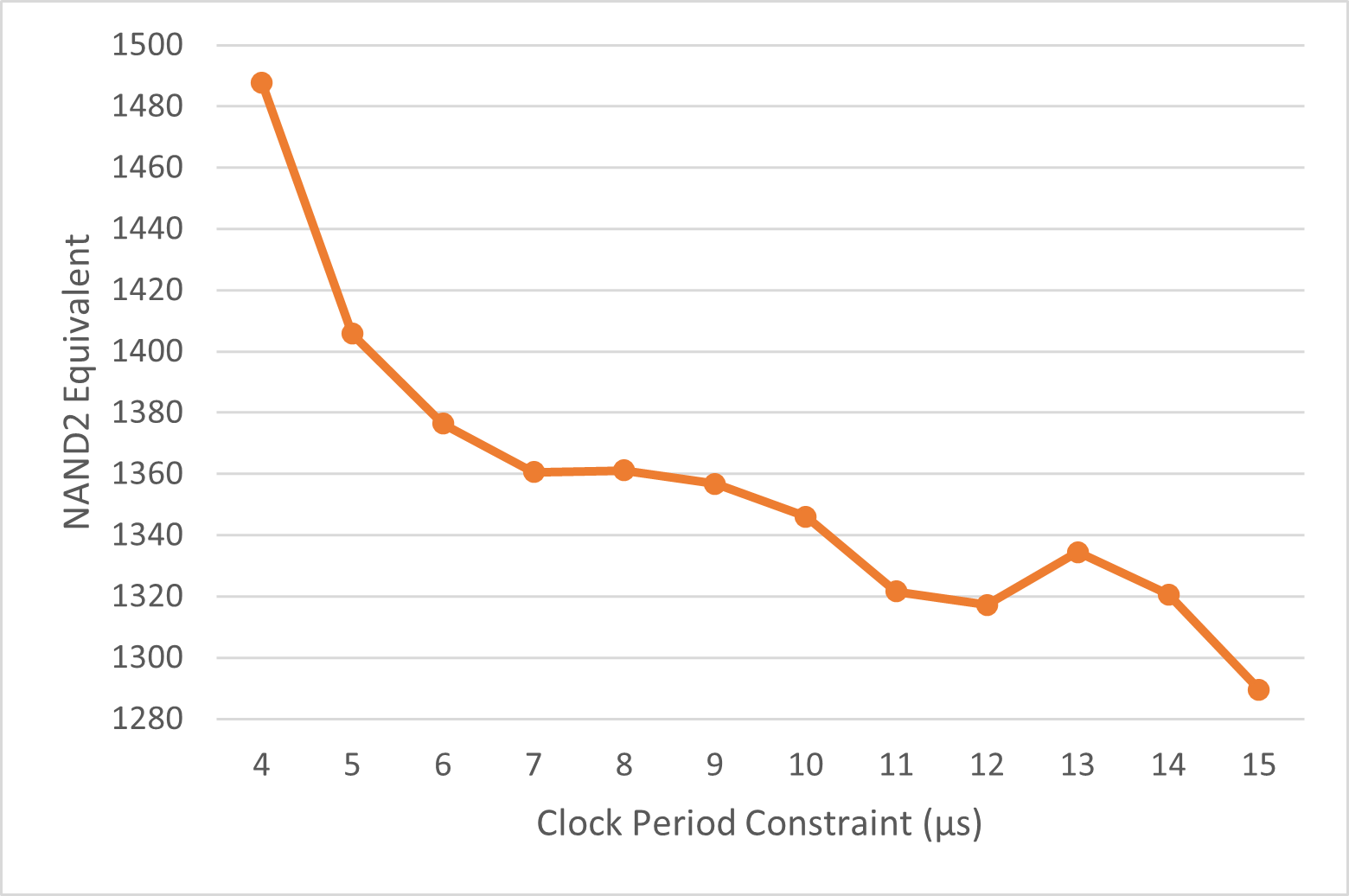}
    \label{NAND2 Comp Model}}
    \caption{Area figure normalised by the area of a minimal strength 2-input NAND gate.}
    \label{fig:area}
\end{figure}

\textbf{Fig.~\ref{fig:area}} depicts the gatecount (NAND2-equivalent) of both TM hardware as a function of the clock period constraint.
We observe that the static (and total) power heavily correlates to the area and gatecount.
The area shows little variation for clock period range, except when the constraint approaches the design limits forcing \textit{Genus} to perform logic duplication and buffer insertion to meet timing.

\begin{figure}
    \centering
    \subfigure[Full Model Energy per Inference]{
    \includegraphics[width=.9\linewidth]{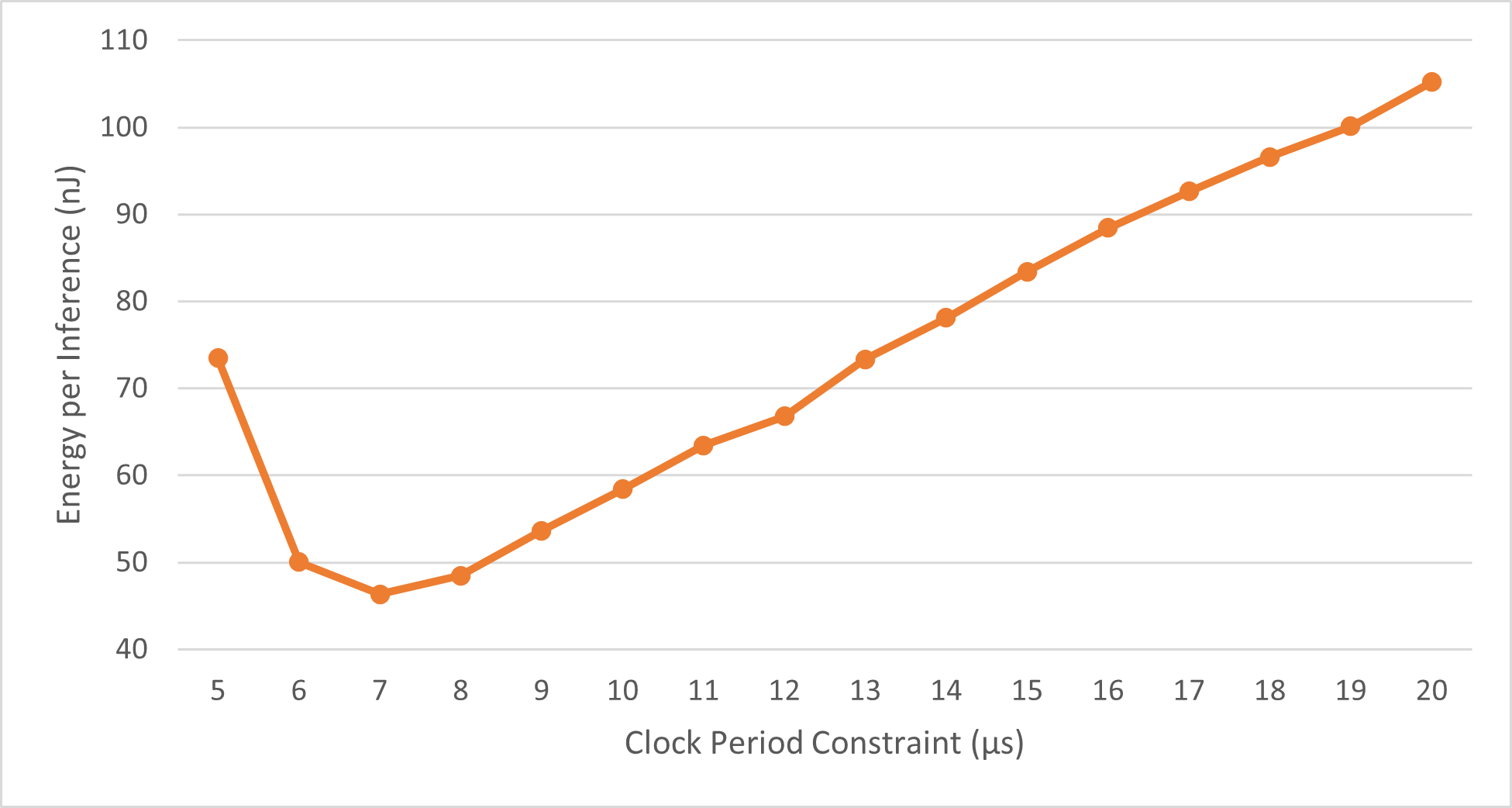}
    \label{Energy Full Model}}
    \quad
    \subfigure[Compact Model Energy per Inference]{
    \includegraphics[width=.9\linewidth]{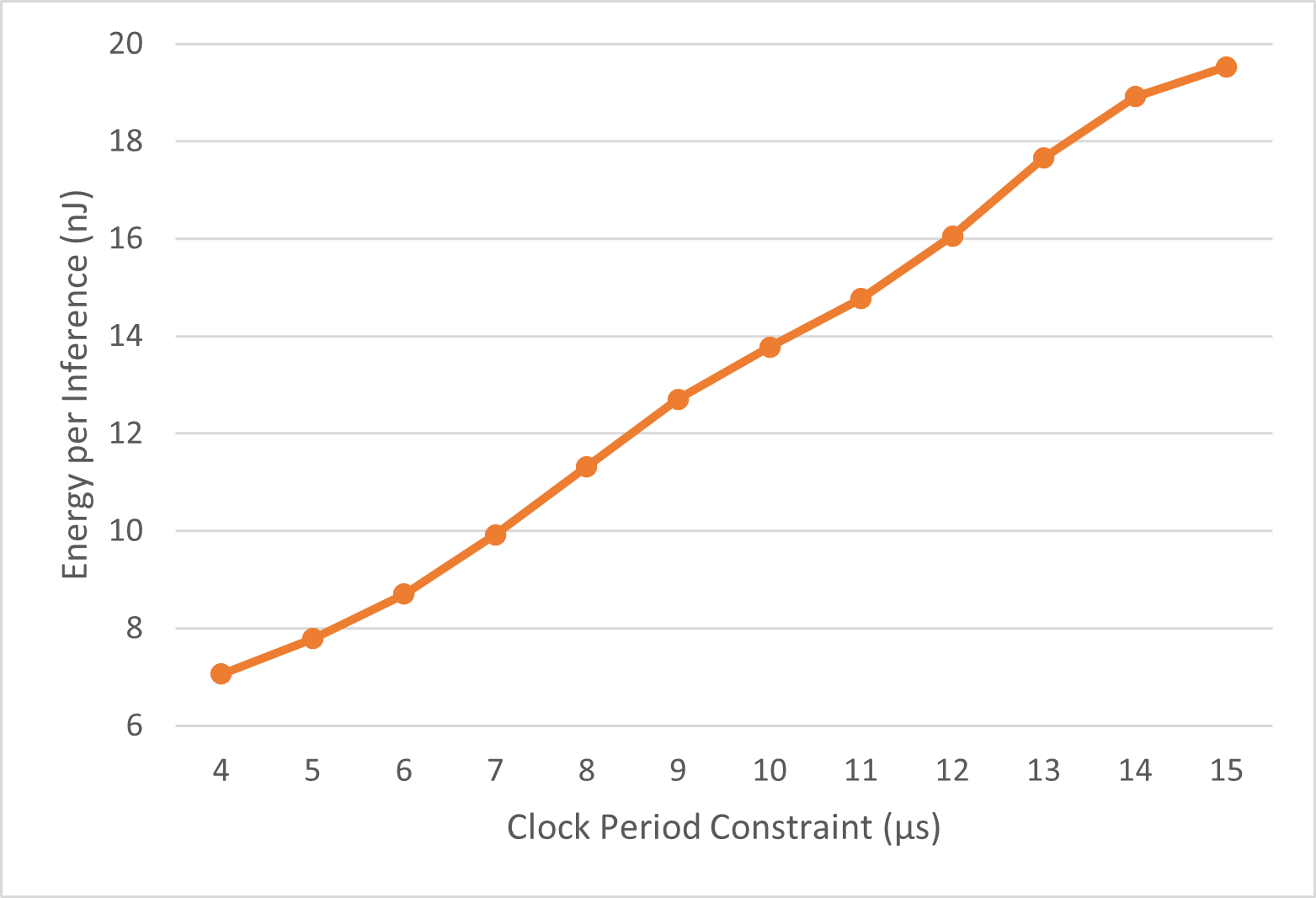}
    \label{Energy Comp Model}}
    \caption{Energy per Inference of Both Models}
    \label{Energy}
\end{figure}

This behaviour indicates that, unlike in conventional CMOS, there is little benefit in reduce the clock period to save power.
In many embedded applications, however, overall power alone is not necessarily the best metric to evaluate the efficiency of a circuit.
It is also important to consider the amount of work required (energy) per operation.
\textbf{Fig. \ref{Energy}} displays the energy per inference for both models as a function of clock cycle constraint.
This results show that since the overall power of the circuit is up to a point hardly affected by the clock constraint, producing a faster circuit yields a better utilisation of the power budget.
For the compact model, the most power efficient circuit is the fastest synthesised circuit.
However, for the full model, the raise in area (consequently power) on constraints bellow \SI{7}{\micro\second} overcomes the gains in operation speed, yielding a less efficient circuit.
These efficient circuits enable power savings when power gating is applied.

\section{Conclusion}
\label{conc}

In this work, we demonstrated the possibility of deploying a complex machine learning accelerator, the Tsetlin Machine, on a flexible integrated circuit substrate. We designed and simulated two prototype chips using Pragmatic’s 600nm IGZO FlexIC process, marking the first realisation of a digital TM on flexible electronics. By using a three-stage pipelined RTL architecture and a strict floorplanning strategy (limiting the core cell density to 10\%), we achieved a robust inference engine with only a three-clock-cycle latency from input to output. The full-scale 8×8mm\textsuperscript{2} chip implemented a TM with nearly 6800 NAND2 equivalent logic gates and attained 98.5\% accuracy on an 8$\times$8-pixel handwritten digit recognition task. The compact 4×4mm\textsuperscript{2} chip, with roughly 1420 NAND2 equivalent gates, realised a streamlined TM model and still achieved 93\% accuracy on the same task. Throughout physical implementation, despite having only two metal interconnect layers, both designs met their timing targets and passed all DRC/LVS verification checks. We ensured power integrity on the flexible chips using dual VDD/GND power pad pairs and on-chip supply routing stripes, which maintained stable operation even under the TM’s switching activity. These results validate that complex, logic-intensive machine learning models can be efficiently mapped onto flexible substrates without significant loss of performance or reliability.

\section{Future Work}
\label{future}

Our work opens several avenues for future research and development. First, the techniques we employed can be extended to next-generation FlexIC processes. Exploring finer feature sizes or processes with additional metal layers could allow larger or more complex TMs (or other ML models) to be implemented on flex, or enable the current TM to handle higher-resolution data. We also plan to investigate on-device learning capabilities – for example, integrating support from online training or adaptive clause pruning directly into the flexible chip. Such features would enable reconfigurable ML algorithms on flexible hardware, which could learn from new data in the field or adjust their complexity to meet power constraints. This would be especially useful in wearable and IoT contexts where the environment may change over time.

Importantly, the fastest clock speeds achieved by our FlexIC prototypes are sufficient for many real-world applications in healthcare and sensing. For instance, wearable biomedical monitors often deal with signals in the few-hundred-Hertz range. A common example is electrocardiography (ECG) for heart monitoring, where a high-resolution signal is typically sampled at around 500Hz to 1kHz. Flexible electronics that have already been applied to on-body ECG analysis – Pillai et al. [12] demonstrate an arrhythmia detection engine on a FlexIC, showing the potential of wearable cardiac monitors. Our TM accelerator, operating in the tens of kHz (e.g. 100kHz in our testing), has several orders of magnitude more processing headroom than needed for real-time ECG processing. This means it can easily handle such analysis of incoming heartbeat data, extracting patterns within each 1ms interval of the signal (which corresponds to 1kHz sampling). As a next step, we plan to evaluate our flexible TM hardware on a real ECG dataset to assess its classification accuracy and latency in detecting arrhythmia or other cardiac events from true physical waveforms. This will help verify that the system’s performance could be translated from image classification to time-series biomedical data.

Another direction for future enhancement is to integrate a lightweight pre-processing module on the flexible chip to handle analogue sensor input. In the case of ECG, this could involve an analogue front-end or an ADC coupled with simple feature extraction logic that converts raw voltage traces into the binary propositional inputs required by the TM. We plan on adding a small on-chip module that takes the streamed ECG signal and produces a set of binary features (for example, threshold crossings, heartbeat interval markers, etc.) which the TM can classify. By including such pre-processing on the same FlexIC, we would have a fully self-contained, end-to-end wearable ML solution – from sensing the analogue biometric signal to producing a classification output – all on a single thin flexible chip. This level of integration would greatly simplify wearable device design (no rigid microcontroller needed) and could improve signal quality (less noise coupling since the analogue path is short). We believe that combining flexible electronics with interpretable machine learning models, such as the TM, is a promising route toward intelligent, skin-mounted health monitors and other novel IoT devices that require both computing and flexibility at the point of use.

\bibliographystyle{ieeetr}
\bibliography{ref/TM, ref/mcculloch_NN_1943, ref/Adrian-TMhardware, ref/biggs_flexARMChip_2021, ref/ozer_malodour_2023, ref/ozer_RISCV_2024}

\end{document}